\documentclass{article}
\usepackage{graphicx}
\usepackage{hyperref}
\usepackage{lineno}
\usepackage[OT1]{fontenc}
\usepackage{amsthm,amsmath,amssymb}
\usepackage[numbers]{natbib}
\usepackage{graphicx}
\usepackage{subfig}
\usepackage{titlesec}
\usepackage{mathrsfs}
\usepackage{multirow}
\usepackage{booktabs}
\RequirePackage{enumitem} 
\RequirePackage{mathtools}
\usepackage[doublespacing]{setspace}
\usepackage{geometry}
\usepackage{subfig}
\usepackage[toc]{appendix}
\numberwithin{equation}{section}
\theoremstyle{plain}

\titlelabel{\thetitle.\quad}
\def\bibindent{1em}

\makeatletter
\renewcommand\@biblabel[1]{} 
\renewenvironment{thebibliography}[1]
{\section*{\refname}%
	\@mkboth{\MakeUppercase\refname}{\MakeUppercase\refname}%
	\list{\@biblabel{\@arabic\c@enumiv}}%
	{\settowidth\labelwidth{\@biblabel{}}%
		\leftmargin\labelwidth
		\advance\leftmargin15pt
		\advance\leftmargin\labelsep
		\setlength\itemindent{-10pt}
		\@openbib@code
		\usecounter{enumiv}%
		\let\p@enumiv\@empty
		\renewcommand\theenumiv{\@arabic\c@enumiv}}%
	\sloppy
	\clubpenalty4000
	\@clubpenalty \clubpenalty
	\widowpenalty4000%
	\sfcode`\.\@m}
{\def\@noitemerr
	{\@latex@warning{Empty `thebibliography' environment}}%
	\endlist}
\renewcommand\newblock{\hskip .11em\@plus.33em\@minus.07em}
\makeatother

\renewcommand{\refname}{References}
\DeclareMathOperator*{\argmin}{argmin}

\begin{document}

\title{\textbf{Design optimisation and post-trial analysis in group sequential stepped-wedge cluster randomised trials}}
\author{\textbf{M. J. Grayling\textsuperscript{1}, D. S. Robertson\textsuperscript{1}, J. M. S. Wason\textsuperscript{1,2}, A. P. Mander\textsuperscript{1}}\\
\small 1. Hub for Trials Methodology Research, MRC Biostatistics Unit, Cambridge, UK, \\ \small 2. Institute of Health and Society, Newcastle University, Newcastle, UK.}
\date{}
\maketitle

\noindent \textbf{Corresponding Author:} M. J. Grayling, MRC Biostatistics Unit, Forvie Site, Robinson Way, Cambridge CB2 0SR, UK; Tel: +44-(0)1223-330300; E-mail: mjg211@cam.ac.uk.\\

\noindent \textbf{Abstract}\\
Recently, methodology was presented to facilitate the incorporation of interim analyses in stepped-wedge (SW) cluster randomised trials (CRTs). Here, we extend this previous discussion. We detail how the stopping boundaries, allocation sequences, and per-cluster per-period sample size of a group sequential SW-CRT can be optimised. We then describe methods by which point estimates, p-values, and confidence intervals, which account for the sequential nature of the design, can be calculated. We demonstrate that optimal sequential designs can reduce the expected required number of measurements under the null hypothesis, compared to the classical design, by up to 30\%, with no cost to the maximal possible required number of measurements. Furthermore, the adjusted analysis procedure almost universally reduces the average bias in the point estimate, and consistently provides a confidence interval with coverage close to the nominal level. In contrast, the coverage of a naive 95\% confidence interval is observed to range between 92 and 98\%. Methodology is now readily available for the efficient design and analysis of group sequential SW-CRTs. In scenarios in which there are substantial ethical or financial reasons to terminate a SW-CRT as soon as possible, trialists should strongly consider a group sequential approach.
\\

\noindent \textbf{Keywords:} Cluster randomized trial; Group sequential; Interim analysis; Stepped-wedge.\\
\\
\section{Introduction}
\label{s:intro}

In a cluster randomised trial (CRT), groups of subjects rather than individuals are randomised to treatment conditions. Today, numerous reasons are recognised for why one may employ a CRT design. Primarily, CRTs aid in the control of contamination between subjects, where participants from the same group would likely influence each other if randomised to different treatments. Furthermore, they can bring increased administrative efficiency, helping to overcome the barriers of contacting and recruiting large numbers of participants, whilst they additionally allow the study of interventions that could not be directed toward a single individual. Unfortunately though, there are also several noted disadvantages to these designs. In particular, since participants from a cluster are likely to be more similar than those from different clusters, there is a potential for correlation between observations from within each cluster. This must be accounted for not only in the analysis, but also at the design stage, as standard sample size calculation formulae are no longer applicable. See, for example, Eldridge and Kerry (2012) \citep{eldridge}, Campbell and Walters (2014) \citep{campbell}, or Hayes and Moulton (2017) \citep{hayes} for further discussions of these issues. 

One sub-category of CRT design that has been the subject of substantial research in recent years is the stepped-wedge (SW) CRT. The key feature of a SW-CRT is that an experimental intervention is introduced across the clusters over a number of time periods using one-directional switching only. That is, once a cluster switches to the experimental intervention in a SW-CRT, it does not revert back to the control intervention for the remainder of the trial. Today, methodology is well established for sample size calculation \citep{hussey,woertman,baio,hooper}, and for data analysis \citep{hemming,wang,kasza}, in fixed sample SW-CRTs.

In addition, techniques for applying classical group sequential trial design methodology to SW-CRTs with interim analyses were recently described \citep{grayling1}. This work followed previous suggestions that early stopping for efficacy may be important in SW-CRTs given the design is regularly associated with scenarios in which the experimental intervention is expected to do more good than harm \citep{dehoop}. Whilst a review of the outcome of completed SW-CRTs led to the proposal that early stopping for futility may be helpful to guard against the over-enthusiastic use of the design \citep{grayling2}. Ultimately, in this work, it was demonstrated that the incorporation of interim analyses in to a SW-CRT could bring substantial efficiency gains in terms of the expected number of required measurements, compared to utilising a classical fixed sample design.

However, one factor left unexamined was how the design of a group sequential SW-CRT could be optimised. Optimal fixed sample SW-CRT design has been a topic of much recent interest \citep{lawrie,girling,thompson,li}, with the same also true of the optimisation of stopping boundaries in conventional individually randomised group sequential trials \citep{wason1,wason2,an}. For both fixed sample SW-CRTs and group sequential trials, design optimisation was demonstrated to provide substantial efficiency gains. We may therefore expect that it would be of great value to establish methodology for optimising the design of group sequential SW-CRTs.

In addition, methodology for post-trial analysis in group sequential SW-CRTs was not considered by Grayling et al. (2017a) \citep{grayling1}. It is well known that if a final analysis is performed on data acquired within a sequential trial, that does not take in to account its sequential nature, then biased point estimates, and p-values and confidence intervals with incorrect coverage will be routinely acquired \citep{jennison}. Accordingly, the lack of available guidance on how to analyse data accrued in a group sequential SW-CRT currently limits the application of such designs in practice.

In this work, we address these problems. Specifically, we first detail how the stopping boundaries, allocation sequences, and per-cluster per-period sample size of a group sequential SW-CRT can be optimised. Following this, we describe an approach for acquiring a median unbiased estimator for the treatment effect, along with methodology for constructing adjusted p-values and confidence intervals. Finally, for two example trial design scenarios, we explore the efficiency gains optimal designs could bring, and examine the performance of the proposed procedures for data analysis.

We now proceed by describing the notation utilised in the rest of the paper. The aforementioned methodology is then presented along with the results of a simulation study. We conclude with a short discussion. 

\section{Methods} \label{s:model}

\subsection{Notation, hypotheses and analysis in group sequential SW-CRTs}\label{gssw}

Define $\mathbb{N}_a=\{k\in\mathbb{N} : k\le a\}$ for any $a\in\mathbb{N}$, and $\mathbb{N}_a^+=\mathbb{N}_a\backslash\{0\}$. Then, we consider a SW-CRT conducted in $C\in\mathbb{N}\backslash\mathbb{N}_1$ clusters over $T\in\mathbb{N}\backslash\mathbb{N}_1$ time periods, with $m\in\mathbb{N}\backslash\mathbb{N}_1$ measurements taken per-cluster per-period, to compare a single experimental intervention to some control. We do not restrict ourselves to cross-sectional designs (in which the measurements are assumed to be taken on different patients in each time period) or cohort designs (in which the measurements are assumed to be taken on the same patients in each time period). However, for simplicity, we focus only on complete block SW-CRTs, where measurements are accrued on every cluster in every time period. Given this, we denote by $\boldsymbol{X}$, $\text{dim}(\boldsymbol{X})=C\times T$, the matrix of binary treatment indicators $X_{c,t}$, where $X_{c,t}=1$ if cluster $c$ receives the intervention in time period $t$, and $X_{c,t}=0$ otherwise.

Denoting the vector of responses available after time period $t\in\mathbb{N}_T^+$ by $\boldsymbol{Y}_{t|m,\boldsymbol{X}}$, we assume that $\boldsymbol{Y}_{t|m,\boldsymbol{X}}\sim N(\boldsymbol{D}_{t|m,\boldsymbol{X}}\boldsymbol{\beta},\boldsymbol{\Sigma}_{t|m,\boldsymbol{X}})$, with $\boldsymbol{\Sigma}_{T|m,\boldsymbol{X}}$ non-singular. Here the subscript $t\in\mathbb{N}_T^+$ indicates we consider the subsections of $\boldsymbol{D}_{T|m,\boldsymbol{X}}$ and $\boldsymbol{\Sigma}_{T|m,\boldsymbol{X}}$ pertaining to the first $t$ time periods only, where $\boldsymbol{D}_{T|m,\boldsymbol{X}}$ and $\boldsymbol{\Sigma}_{T|m,\boldsymbol{X}}$ are designated design and covariance matrices for the vector of responses $\boldsymbol{Y}_{T|m,\boldsymbol{X}}$. Furthermore, $\boldsymbol{\beta}=(\beta_1,\dots,\beta_p)^\top$ is a specified vector of fixed effects. Note that we have explicitly stated the distribution conditional on the values of $m$ and $\boldsymbol{X}$, since these will later be treated as variables to be determined.

We assume that we would like to make inference about the final element of $\boldsymbol{\beta}$, $\beta_p$, which we denote for simplicity by $\tau$. Formally, we test the one-sided hypothesis $H_0 : \tau\le0$, against $H_1 : \tau>0$.

A group sequential SW-CRT can then be completely defined given
\begin{itemize}
	\item A set of integers $\mathscr{T}=\{t_1,\dots,t_{|\mathscr{T}|}\}$, with $t_i\in\mathbb{N}_T^+$ for $i\in\mathbb{N}_{|\mathscr{T}|}^+$, $t_i>t_{i-1}$ for $i\in\mathbb{N}_{|\mathscr{T}|}\backslash\mathbb{N}_1$, and $t_{|\mathscr{T}|} =T$. 
	\item Futility and efficacy boundaries $\boldsymbol{f}=(f_1,\dots,f_{|\mathscr{T}|})^\top$ and $\boldsymbol{e}=(e_1,\dots,e_{|\mathscr{T}|})^\top$, with $(\boldsymbol{f},\boldsymbol{e})\in\mathscr{B}_{|\mathscr{T}|}=\{(\boldsymbol{a},\boldsymbol{b})\in\mathbb{R}^{|\mathscr{T}|}\times\mathbb{R}^{|\mathscr{T}|} : \forall i\in\mathbb{N}_{|\mathscr{T}|-1}^+\ a_i<b_i,\ a_{|\mathscr{T}|}=b_{|\mathscr{T}|}\}$.
\end{itemize}

With this, $|\mathscr{T}|$ analyses are carried out in total, with analysis $i\in\mathbb{N}_{|\mathscr{T}|}^+$ conducted after time period $t_i\in\mathbb{N}_T^+$. Explicitly, for each $i\in\mathbb{N}_{|\mathscr{T}|}^+$ we compute an estimate for $\boldsymbol{\beta}$, $\hat{\boldsymbol{\beta}}_{i|m,\boldsymbol{X}}$, using generalised least squares
\begin{align}
\hat{\boldsymbol{\beta}}_{i|m,\boldsymbol{X}} &= (\hat{\beta}_{1,i|m,\boldsymbol{X}},\dots,\hat{\beta}_{p,i|m,\boldsymbol{X}})^\top,\nonumber\\
&= (\boldsymbol{D}_{t_i|m,\boldsymbol{X}}^\top \boldsymbol{\Sigma}_{t_i|m,\boldsymbol{X}}^{-1} \boldsymbol{D}_{t_i|m,\boldsymbol{X}})^{-1} \boldsymbol{D}_{t_i|m,\boldsymbol{X}}^\top \boldsymbol{\Sigma}_{t_i|m,\boldsymbol{X}}^{-1} \boldsymbol{Y}_{t_i|m,\boldsymbol{X}}.\label{analysis}
\end{align}
From this, we use the estimate of $\tau$,  $\hat{\tau}_{i|m,\boldsymbol{X}}=\hat{\beta}_{p,i|m,\boldsymbol{X}}$, to form the following standardised test statistic
\[ Z_{i|m,\boldsymbol{X}}=\frac{\hat{\tau}_{i|m,\boldsymbol{X}}}{\{\text{Var}(\hat{\tau}_{i|m,\boldsymbol{X}})\}^{1/2}} = \hat{\tau}_{i|m,\boldsymbol{X}} \{(\boldsymbol{D}_{t_i|m,\boldsymbol{X}}^\top \boldsymbol{\Sigma}_{t_i|m,\boldsymbol{X}}^{-1} \boldsymbol{D}_{t_i|m,\boldsymbol{X}})_{[p,p]}^{-1} \}^{-1/2} = \hat{\tau}_{i|m,\boldsymbol{X}} I_{i|m,\boldsymbol{X}}^{1/2}. \]

Note that for this to be possible (i.e., for $\tau$ to be estimable at every interim analysis), it is important that at least one cluster receives the intervention in one of the time periods $t\in\mathbb{N}_{t_1}^+$. This can be verified using the matrix $\boldsymbol{X}$ by checking $\sum_{c\in\mathbb{N}_C^+}\sum_{t\in\mathbb{N}_{t_1}^+}X_{c,t}>0$.

An algorithm for conducting a group sequential SW-CRT can then be defined as follows

\begin{enumerate}
	\item Set $t=1$.
	\item Conduct period $t$ of the trial, allocating treatments according to the values $X_{c,t}$ for $c\in\mathbb{N}_C^+$, and accruing $m$ measurements in each cluster.
	\item If $t\notin\mathscr{T}$ set $t=t+1$ and return to 2. Otherwise, if $t\in\mathscr{T}$, $t=t_i$ say, compute $Z_{i|m,\boldsymbol{X}}$. Then
	\begin{itemize}
		\item If $Z_{i|m,\boldsymbol{X}}\le f_i$ stop the trial, and do not reject $H_0$.
		\item If $Z_{i|m,\boldsymbol{X}}> e_i$ stop the trial, and reject $H_0$.
		\item If $f_i < Z_{i|m,\boldsymbol{X}} \le e_i$, set $t=t+1$ and return to 2.
	\end{itemize}
\end{enumerate}

Here, the trial is guaranteed to terminate, with a decision also made on $H_0$, by the end of time period $T$ because of the stipulation that $f_{|\mathscr{T}|}=e_{|\mathscr{T}|}$.

Importantly, $(Z_{1|m,\boldsymbol{X}},\dots,Z_{|\mathscr{T}||m,\boldsymbol{X}})^\top$ is multivariate normal with
\begin{align*}
\mathbb{E}(Z_{i|m,\boldsymbol{X}}) &= \tau I_{i|m,\boldsymbol{X}}^{1/2}, \qquad i\in\mathbb{N}_{|\mathscr{T}|}^+,\\
\text{Cov}(Z_{i|m,\boldsymbol{X}},Z_{j|m,\boldsymbol{X}}) &= (I_{i|m,\boldsymbol{X}}/I_{j|m,\boldsymbol{X}})^{1/2}, \qquad i,j\in\mathbb{N}_{|\mathscr{T}|}^+,\ i\le j,
\end{align*}
and so multivariate normal integration can be used to evaluate the statistical operating characteristics of the design \citep{jennison}.

To see this clearly, define the random variable $\Gamma\in\mathbb{N}_{|\mathscr{T}|}^+$ as the analysis at which the trial is terminated, and $\Psi\in\mathbb{N}_1$ as the random variable with $\Psi=1$ if $H_0$ is rejected, and $\Psi=0$ otherwise. The possible outcomes to the trial are then categorised by the pair $\{\Psi,\Gamma\}$, and for any $\{\psi,\gamma\}\in\mathbb{N}_1\times\mathbb{N}_{|\mathscr{T}|}^+$
\begin{small}
	\begin{equation}\label{probab}
	\mathbb{P}(\Gamma=\gamma,\Psi=\psi | \tau,\boldsymbol{f},\boldsymbol{e},m,\boldsymbol{X}) = \int_{\text{l}(1,\gamma,\psi,\boldsymbol{f},\boldsymbol{e})}^{\text{u}(1,\gamma,\psi,\boldsymbol{f},\boldsymbol{e})}\dots\int_{\text{l}(\gamma,\gamma,\psi,\boldsymbol{f},\boldsymbol{e})}^{\text{u}(\gamma,\gamma,\psi,\boldsymbol{f},\boldsymbol{e})}\phi(\boldsymbol{y},\tau\boldsymbol{I}_{\gamma|m,\boldsymbol{X}}^{1/2},\boldsymbol{\Lambda}_{\gamma|m,\boldsymbol{X}})\mathrm{d}y_\gamma\dots\mathrm{d}y_1,
	\end{equation}
\end{small}
where 
\begin{itemize}
	\item $\phi\{\boldsymbol{y},\boldsymbol{\mu},\boldsymbol{\Lambda}\}$ is the probability density function of a multivariate normal distribution with mean \(\boldsymbol{\mu}\) and covariance matrix \(\boldsymbol{\Lambda}\), evaluated at vector $\boldsymbol{y}$,
	\item $\boldsymbol{I}_{\gamma|m,\boldsymbol{X}}^{1/2} = (I_{1|m,\boldsymbol{X}}^{1/2},\dots,I_{\gamma|m,\boldsymbol{X}}^{1/2})^\top$.
	\item $\text{l}$ and $\text{u}$ are functions that provide the lower and upper integration limits for each integral. Explicitly, for any $(\boldsymbol{f},\boldsymbol{e})\in\mathscr{B}_{|\mathscr{T}|}$
	\begin{align*}
	\text{l}(i,\gamma,\psi,\boldsymbol{f},\boldsymbol{e}) = \begin{cases} f_i &:\ i<\gamma, \\ e_i &:\ i=\gamma,\ \psi=1, \\ -\infty  &:\ i=\gamma, \psi=0, \end{cases}\qquad 
	\text{u}(i,\gamma,\psi,\boldsymbol{f},\boldsymbol{e}) = \begin{cases} e_i &:\ i<\gamma, \\ \infty &:\ i=\gamma,\ \psi=1, \\ f_i  &:\ i=\gamma, \psi=0. \end{cases}
	\end{align*}
	\item $\boldsymbol{\Lambda}_{\gamma|m,\boldsymbol{X}}$ is the covariance matrix of the standardized test statistics across the interim analyses $1,\dots,\gamma$
	\[ \boldsymbol{\Lambda}_{\gamma|m,\boldsymbol{X}} =
	\begin{pmatrix}
	\text{Cov}(Z_{1|m,\boldsymbol{X}},Z_{1|m,\boldsymbol{X}}) & \dots & \text{Cov}(Z_{1|m,\boldsymbol{X}},Z_{\gamma|m,\boldsymbol{X}}) \\[0.3em]
	\vdots & \ddots & \vdots \\[0.3em]
	\text{Cov}(Z_{\gamma|m,\boldsymbol{X}},Z_{1|m,\boldsymbol{X}}) & \dots & \text{Cov}(Z_{\gamma|m,\boldsymbol{X}},Z_{\gamma|m,\boldsymbol{X}})
	\end{pmatrix}. \]
\end{itemize}
	
With this, the probability that $H_0$ is rejected, and the expected number of required measurements (ENM), can be computed as

\begin{align*}
P(\tau\mid\boldsymbol{f},\boldsymbol{e},m,\boldsymbol{X}) &= \sum_{\gamma\in\mathbb{N}_{|\mathscr{T}|}^+}\mathbb{P}(\Gamma=\gamma,\Psi=1 | \tau,\boldsymbol{f}_\gamma,\boldsymbol{e}_\gamma,m,\boldsymbol{X}),\\
ENM(\tau\mid\boldsymbol{f},\boldsymbol{e},m,\boldsymbol{X}) &= \sum_{\gamma\in\mathbb{N}_{|\mathscr{T}|}^+}\sum_{\psi\in\mathbb{N}_1}mCt_{\gamma}\mathbb{P}(\Gamma=\gamma,\Psi=\psi | \tau,\boldsymbol{f}_\gamma,\boldsymbol{e}_\gamma,m,\boldsymbol{X}).
\end{align*}

\subsection{Optimal group sequential designs}

Section~\ref{gssw} describes a method for conducting a group sequential SW-CRT given choices for several design parameters. However, at the design stage, we require a method for choosing values for these parameters to provide desired operating characteristics. One simple approach to this, after designating a value for $\boldsymbol{X}$, is the error spending method, as discussed in Grayling et al. (2017a) \citep{grayling1}. Here, we consider instead how $\boldsymbol{X}$, $\boldsymbol{f}$, $\boldsymbol{e}$, and either $C$ or $m$, can be optimised. 

For this, much depends on whether the choice is to fix $C$ or $m$. In the former case, the matrix $\boldsymbol{X}$ can be readily optimised, as will be described shortly. In the latter case, optimisation of $\boldsymbol{X}$ is complex since its row dimension is variable in the search for the required value of $C$. In this case, $\boldsymbol{X}$ should be specified implicitly through rules on what proportion of clusters should switch to the intervention in each time period, reducing the problem to optimising only $\boldsymbol{f}$, $\boldsymbol{e}$, and $C$. Then, having determined an appropriate value for $C$, one can employ the approach for fixed $C$ to optimise $\boldsymbol{X}$ if desired.

Thus, from here, we focus on the problem of simultaneously optimising $\boldsymbol{X}$, $\boldsymbol{f}$, $\boldsymbol{e}$, and $m$, when $C$ has been fixed. To this end, define the vector $\boldsymbol{S}=(S_1,\dots,S_C)^\top\in\mathscr{S}_{\mathscr{T}}$ of switching times, where
\[ \mathscr{S}_{\mathscr{T}}=\left\{\boldsymbol{s}\in\{\mathbb{N}_{T+1}^+\}^{C} : \min_{c\in\mathbb{N}_C^+} s_c\le t_1,\ \forall t\in\mathbb{N}_{T}^+ \prod_{c\in\mathbb{N}_C^+}\mathbb{I}(s_c=t)=0 \right\}. \]
That is, $X_{c,t}=\mathbb{I}(t \ge S_c)$ for $(c,t)\in\mathbb{N}_C^+\times\mathbb{N}_T^+$. Here, we have used $S_c=T+1$ to indicate that cluster $c$ does not receive the intervention during the trial, whilst the additional constraints ensure $\tau$ is estimable at every interim analysis, and that at least two distinct sequences of allocations are used across the clusters. We will use the notation $\boldsymbol{X}(\boldsymbol{S})$ to indicate the matrix $\boldsymbol{X}$ implied by a particular choice for $\boldsymbol{S}$.

We then consider an optimal design framework where we wish to identify the solution to a problem which can be cast in the following form

\begin{equation*}
\argmin_{\{(\boldsymbol{f},\boldsymbol{e}),m,\boldsymbol{S}\}\in\mathscr{B}_{|\mathscr{S}|}\times(\mathbb{N}\backslash\mathbb{N}_1)\times\mathscr{S}_{\mathscr{T}}} O(\boldsymbol{f},\boldsymbol{e},m,\boldsymbol{S}),
\end{equation*}
subject to
\begin{equation*}
P\{0|\boldsymbol{f},\boldsymbol{e},m,\boldsymbol{X}(\boldsymbol{S})\}\le\alpha,\qquad
P\{\delta|\boldsymbol{f},\boldsymbol{e},m,\boldsymbol{X}(\boldsymbol{S})\}\ge1 - \beta.
\end{equation*}
Here, the constraints are interpreted as controlling the type-I error-rate to some $\alpha\in(0,1)$ when $\tau=0$, and to have power of at least $1-\beta\in(0,1)$ when $\tau=\delta>0$.

In practice, the objective function $O(\cdot)$ could be specified to be as complex as desired. For example, as discussed in Grayling et al. (2017c) \citep{grayling3}, it could penalise designs which roll-out the intervention beyond a certain speed, or impose a cost on designs where clusters do not receive the experimental intervention. Here though, adapting previous relevant works on optimal group sequential trial design \citep{wason1,wason2}, we consider
\begin{equation*}
O(\boldsymbol{f},\boldsymbol{e},m,\boldsymbol{S})=w_1 ENM\{0|\boldsymbol{f},\boldsymbol{e},m,\boldsymbol{X}(\boldsymbol{S})\}+w_2 ENM\{\delta|\boldsymbol{f},\boldsymbol{e},m,\boldsymbol{X}(\boldsymbol{S})\}+w_3 mCT.
\end{equation*}
With this, the $w_i\in\mathbb{R}^+\cup\{0\}$ are weights indicating the relative strengths of the desires to minimise the ENM under the null, the ENM under the alternative hypothesis, and the maximal required number of observations. Later, we will use the notation $\boldsymbol{w}=(w_1,w_2,w_3)$.

Unfortunately, the complexity of the constraints on the trial's operating characteristics means they cannot easily be incorporated in to a conventional numerical optimisation routine that we may seek to use to solve the above problem. Therefore, following Wason and Jaki (2012) \citep{wason1} and Wason et al. (2012) \citep{wason2}, we translate our problem to be
\begin{small}
	\begin{equation}\label{problem}
	\begin{split}
	\argmin_{\{(\boldsymbol{f},\boldsymbol{e}),m,\boldsymbol{S}\}\in\mathscr{B}_{|\mathscr{S}|}\times(\mathbb{N}\backslash\mathbb{N}_1)\times\mathscr{S}_{\mathscr{T}}} & O(\boldsymbol{f},\boldsymbol{e},m,\boldsymbol{S})+M_{SW}\left( \mathbb{I}[P\{0|\boldsymbol{f},\boldsymbol{e},m,\boldsymbol{X}(\boldsymbol{S})\}>\alpha]\left[\frac{P\{0|\boldsymbol{f},\boldsymbol{e},m,\boldsymbol{X}(\boldsymbol{S})\}-\alpha}{\alpha}\right] \right.\\
	&\qquad\qquad+\left. \mathbb{I}[1-P\{\delta|\boldsymbol{f},\boldsymbol{e},m,\boldsymbol{X}(\boldsymbol{S})\}>\beta]\left[\frac{1-P\{\delta|\boldsymbol{f},\boldsymbol{e},m,\boldsymbol{X}(\boldsymbol{S})\}-\beta}{\beta}\right] \right).
	\end{split}
	\end{equation}
\end{small}
Here $M_{SW}$ is the number of observations required by a corresponding fixed-sample near-balanced SW-CRT (i.e., a design with as near-equal as possible numbers of clusters switching to the intervention in each time period). The final factor is thus a penalty for designs that do not meet the desired operating characteristics, to assist in converging towards a solution meeting these requirements.

The final challenge therefore is to identify the solution of the problem~(\ref{problem}). For this, a numerical optimisation routine capable of handling the fact that $m$ and the $S_c$ are discrete, but the components of $\boldsymbol{f}$ and $\boldsymbol{e}$ are continuous, is required. This algorithm must also be capable of handling the many local optima the search space will likely contain. One such routine, available in \texttt{R}, is the \texttt{CEoptim} function from the package of the same name \citep{benham}. We utilise this for all examples given here, and in the Appendix we elaborate on our choices for the algorithms initialisation. 

Our full procedure for optimising the design of a group sequential SW-CRT has then been prescribed. In Section~\ref{optdes} we present numerous such optimal designs.

\subsection{Post trial analysis: Point estimates, p-values, and confidence intervals}\label{analysissec}

In this section, we turn our attention to the final analysis of data accrued in a completed group sequential SW-CRT. For notational simplicity, we therefore suppose that the trial was conducted using design $\mathscr{D}=\{\mathscr{T},\boldsymbol{f}, \boldsymbol{e},m,\boldsymbol{X}\}$ (i.e., these factors are all fixed and known in what follows). Additionally, we assume that the trial terminated with $\{\Gamma,Z_{\Gamma|m,\boldsymbol{X}}\}=\{\gamma,z_{\gamma}\}\in\Xi_{\mathscr{D}} = \{\{i,z_i\}\in\mathscr{T}\times\mathbb{R} : z_i\notin(f_i,e_i]\}$.

With this, a simple method for specifying a point estimate for the treatment effect $\tau$, is to use the maximum likelihood estimate (MLE) $\hat{\tau}_{\text{N}} = \hat{\tau}_{\gamma|m,\boldsymbol{X}}=\hat{\beta}_{p,\gamma|m,\boldsymbol{X}}$,
as defined by Equation~(\ref{analysis}). Additionally, one could designate a p-value as $p_{\text{N}}=1 - \Phi(z_\gamma)$, where $\Phi(z)$ is the cumulative density function of the standard normal distribution. Finally, a one-sided confidence interval for $\tau$, could be taken as
$C_{\text{N}}=(\hat{\tau}_{\text{LCI,N}},\infty)$ for $\hat{\tau}_{\text{LCI,N}}=\hat{\tau}_{\text{N}}-\Phi^{-1}(1-\alpha)I_{\gamma|m,\boldsymbol{X}}^{-1/2}$.

The problem with these methods however is that they do not take into account the sequential nature of the design. As was noted in Section 1, such naive analysis procedures, as the subscript N is meant to indicate, will routinely result in biased point estimates, and p-values and confidence intervals with incorrect coverage. For example, Tsiatis et al. (1984) \citep{tsiatis} demonstrated that the coverage of a naive 90\% confidence interval at the end of a four-stage group sequential trial could be as low as 0.846.

Accordingly, methodology for performing an adjusted analysis is required. Unfortunately, there is no method that is optimal for analysis following a sequential trial in all situations. Whitehead (1986) \citep{whitehead} proposed one of the first methods for analysing data accrued in a group sequential trial, suggesting that the point estimate be taken as the solution to a simple equation involving the computation of the expected bias in the estimated treatment effect. This estimator was demonstrated to typically have extremely small bias. However, its evaluation is complex and it provides no easy means of nominating an associated confidence interval with the desired coverage. Emerson and Fleming (1990) \citep{emerson} suggested use of a uniform minimum variance unbiased estimator, formed through Rao-Blackwellisation of the unbiased estimate of $\tau$ from the first stage data. Unfortunately, this estimator was found to have large variance.
 
Consequently, we instead present an approach that is both easy to apply, and generally has good performance. We use the so-called stage-wise ordering in a group sequential design to designate p-values and confidence intervals with exact coverage probabilities, and to specify a median unbiased point estimate for $\tau$. These computations will require evaluation only of integrals of the form in Equation~(\ref{probab}).

The stage-wise ordering was first proposed by Armitage (1957) \citep{armitage}, and has subsequently been utilised in many papers on individually randomised group sequential trials (see, for example, Siegmund (1978) \citep{siegmund} and Tsiatis et al. (1984) \citep{tsiatis}). For any $\{i,z_i\},\{j,z_j\}\in\Xi_{\mathscr{D}}$, the stage-wise ordering considers $\{i,z_i\}$ to be more extreme than $\{j,z_j\}$ if $i<j$ or if $i=j$ and $z_i\ge z_j$. This ordering can then be used to derive our adjusted p-values and confidence intervals, as well as the median unbiased point estimate. Specifically, define
\begin{align*}
   E(\tau\mid i,z_{i},\boldsymbol{f},\boldsymbol{e},m,\boldsymbol{X}) &= \sum_{j\in\mathbb{N}_{i-1}^+} \mathbb{P}(\Gamma=j,\Psi=1 | \tau,\boldsymbol{f}_j,\boldsymbol{e}_j,m,\boldsymbol{X}),\\ &\qquad +\mathbb{P}\{\Gamma=i,\Psi=1 | \tau,\boldsymbol{f}_i,(\boldsymbol{e}_{i-1}^\top,z_i)^\top,m,\boldsymbol{X}\},
\end{align*}
which is the probability, for the given stage-wise ordering, of seeing a result more extreme than $(i,z_i)$ for a particular value of $\tau$.

Using this, an overall $p$-value for $H_0$ can be taken as $p_{\text{SO}} = E(0\mid \gamma,z_{\gamma},\boldsymbol{f},\boldsymbol{e},m,X)$. Similarly, the associated adjusted confidence interval is $C_{\text{SO}} = (\hat{\tau}_{\text{LCI,SO}},\infty)$, where $\hat{\tau}_{\text{LCI,SO}}$ is the solution of  $\alpha = E(\hat{\tau}_{\text{LCI,SO}}\mid \gamma,z_{\gamma},\boldsymbol{f},\boldsymbol{e},m,\boldsymbol{X})$. Finally, a median unbiased estimate can be designated as the solution $\hat{\tau}_{\text{SO}}$ to $0.5 = E(\hat{\tau}_{\text{SO}}\mid \gamma,z_{\gamma},\boldsymbol{f},\boldsymbol{e},m,\boldsymbol{X})$.

Both $\hat{\tau}_{\text{LCI,SO}}$ and $\hat{\tau}_{\text{SO}}$ can be determined using a simple one-dimensional root finding algorithm, and thus we have established a potential means of performing an improved final analysis on data in group sequential SW-CRTs.

To examine the statistical performance of these analysis procedures, we simulate $R=10^5$ trials for several optimised group sequential designs $\mathscr{D}$ and values for $\tau$. The details of the data degeneration procedure in the simulation study are described in the Appendix. From this, for each considered design we compute, for $r\in\mathbb{N}_R^+$, $C_{x,r}(\tau|\mathscr{D})$ and $\hat{\tau}_{x,r}(\tau|\mathscr{D})$ for $x\in\{\text{N},\text{SO}\}$. Here, $\hat{\tau}_{x,r}(\tau|\mathscr{D})$ is, in a trial using design $\mathscr{D}$, the point estimate at the final analysis for simulation $r$ using estimator $x$, when the true treatment effect is $\tau$, with an analogous definition holding for $C_{x,r}(\tau)$.

The performance of the two point estimation procedures is then assessed using the average bias and average root mean square error (RMSE), computed as
\begin{equation*}
b(\tau\mid x,\mathscr{D}) = R^{-1}\sum_{r\in\mathbb{N}_R^+}\{\hat{\tau}_{x,r}(\tau|\mathscr{D})-\tau\},\qquad
RMSE(\tau\mid x,\mathscr{D}) = \sqrt{R^{-1}\sum_{r\in\mathbb{N}_R^+}\{\hat{\tau}_{x,r}(\tau|\mathscr{D})-\tau\}^2}.\\
\end{equation*}
Similarly, the confidence intervals are compared using their respective coverages, given by
\begin{equation*}
COV(\theta\mid x,\mathscr{D}) = R^{-1}\sum_{r\in\mathbb{N}_R^+}\mathbb{I}\{\tau\in C_{x,r}(\tau|\mathscr{D})\}.
\end{equation*}

\subsection{Examples}\label{example}

We follow Grayling et al. (2017a) \citep{grayling1} and consider two example trial design scenarios (TDSs). In both, we assume a cross-sectional design, with analysis conducted utilising the following linear mixed model, as proposed by Hussey and Hughes (2007) \citep{hussey}
\[ y_{ijk}=\mu+\pi_j+\tau X_{ij}+c_i+\epsilon_{ijk}, \]
for $i\in\mathbb{N}_C^+$, $j\in\mathbb{N}_T^+$, and $k\in\mathbb{N}_m^+$. Here
\begin{itemize}
	\item $y_{ijk}$ is the response of the $k$th individual in the $i$th cluster and $j$th time period.
	\item $\mu$ is an intercept term.
	\item $\pi_j$ is a fixed effect for the $j$th period (with $\pi_1=0$ for identifiability).
	\item $c_i \sim N(0,\sigma_c^2)$ is a random effect for the $i$th cluster.
	\item $e_{ijk} \sim N(0,\sigma_e^2)$ is the residual error.
\end{itemize}

Now, for TDS1, we consider the SW-CRT reported by Bashour et al. (2013) \citep{bashour}, on the effect of training doctors in communication skills on women's satisfaction with doctor-woman relationship during labour and delivery. In this trial, a balanced four cluster five period design was used, and the final analysis estimated that $\sigma_c^2=0.02$ and $\sigma_e^2=0.51$. Based on this, a sample size of 70 patients per-cluster per-period would have been required for the trial's desired operating characteristics ($\alpha=0.05$, $\beta=0.1$, and $\delta=0.2$), implying a total sample requirement of 1400 subjects. Thus, in TDS1 we fix $C=4$, $T=5$, $\sigma_c^2=0.02$, $\sigma_e^2=0.51$, $\alpha=0.05$, $\beta=0.1$, $\delta=0.2$, and $M_{SW}=1400$.

For TDS2, we base our design parameters on the average characteristics of completed SW-CRTs as reported in Grayling et al. (2017a) \citep{grayling1}. We set $C=20$ and $T =9$, to correspond to the median values used in-practice. Next, we suppose $\alpha=0.05$, $\beta=0.2$,
and choose $\sigma_e^2=1$, $\sigma_c^2=1/9$, to provide a more moderate value for the intra-cluster correlation compared
to TDS1. In a near-balanced design (i.e., one where three clusters switch to the intervention in the second through fifth periods, and two clusters in each of the remaining time periods), a design with $m=7$ and $\delta=0.24$ provides the desired power and type-I error-rate. This will therefore be our designated value of $\delta$, and $M_{SW}=1260$.

Note that in these scenarios, $\boldsymbol{D}_{T|m,\boldsymbol{X}}$ and $\boldsymbol{\Sigma}_{T|m,\boldsymbol{X}}$ can be specified easily for any $m$ and $\boldsymbol{X}$. Moreover, recalling that $C$ and $T$ are designated for any $\boldsymbol{X}$ by the relationship $\text{dim}(\boldsymbol{X})=C\times T$, we can then easily determine $I_{t|m,\boldsymbol{X}}$ using the result of Hussey and Hughes (2007) \citep{hussey}
\[ I_{t|m,\boldsymbol{X}} = \frac{(\sigma^2+t\sigma_c^2)(CU-W) + \sigma_c^2(U^2-CV)}{C\sigma^2(\sigma^2+t\sigma_c^2)}, \]
for $\sigma^2=\sigma_e^2/m$, and where
\[ U=\sum_{c\in\mathbb{N}_C^+}\sum_{j\in\mathbb{N}_t^+}X_{c,j},\qquad V=\sum_{c\in\mathbb{N}_C^+}\left(\sum_{j\in\mathbb{N}_t^+}X_{c,j}\right)^2,\qquad W=\sum_{j\in\mathbb{N}_t^+}\left(\sum_{c\in\mathbb{N}_C^+}X_{c,j}\right)^2.  \]

Code to replicate our determination of optimal designs, and to explore the performance of the considered analysis procedures, is available from https://github.com/mjg211/article\_code.

\section{Results}\label{results}

\subsection{Optimal designs}\label{optdes}

For the TDSs described in Section~\ref{example}, a numerical search was performed to identify the optimal values of $\boldsymbol{e}$, $\boldsymbol{f}$, $m$, and $\boldsymbol{X}$ for $\boldsymbol{w}\in\{(1/3,1/3,1/3),\ (1/2,1/2,0),\ (0,1/2,1/2)\}$, with $\mathscr{T}=\{3,5\}$ in TDS1 and $\mathscr{T}=\{3,6,9\}$ in TDS2. The determined optimal designs, and a summary of their statistical operating characteristics, are presented in Table~\ref{tab1}. Their type-I error-rates and power are omitted as each was verified to be the desired level.

\begin{footnotesize}
	\begin{table}[htbp]
		\renewcommand{\arraystretch}{1.2}
		\begin{center}
			\begin{footnotesize}
				\begin{tabular}{rrrrrrrr}
					\toprule
					TDS & $\boldsymbol{w}$ & $m$ & $\boldsymbol{S}^\top$ & $\boldsymbol{f}^\top$ & $\boldsymbol{e}^\top$ & $ENM(0)$ & $ENM(\delta)$ \\
					\midrule
					\multirow{3}{*}{TDS1} & $(1/3, 1/3, 1/3)$ & 69 & (1,2,3,5) & (0.41,1.66) & (2.27,1.66) & 1010.0 & 1073.7 \\
					& $(1/2, 0, 1/2)$ & 70 & (1,2,3,5) & (0.68,1.60) & (2.95,1.60) & 978.6 & 1219.0 \\
					& $(0, 1/2, 1/2)$ & 69 & (1,2,3,5) & (-5.05,1.71) & (2.12,1.71) & 1370.7 & 1055.8 \\
					\midrule
					\multirow{6}{*}{TDS2} & \multirow{2}{*}{$(1/3, 1/3, 1/3)$} & \multirow{2}{*}{7} & $(1,1,1,2,2,2,3,3,4,4,$ & \multirow{2}{*}{(-0.07,0.67,1.65)} & \multirow{2}{*}{(2.64,2.14,1.65)} & \multirow{2}{*}{725.5} & \multirow{2}{*}{923.2} \\
					& & & $5,5,6,6,6,8,8,8,9,10)$ & & & & \\
					& \multirow{2}{*}{$(1/2, 0, 1/2)$} & \multirow{2}{*}{7} & $(1,1,1,1,2,2,2,3,3,3,$ & \multirow{2}{*}{(0.04,0.77,1.58)} & \multirow{2}{*}{(14.41,12.93,1.58)} & \multirow{2}{*}{705.7} &  \multirow{2}{*}{1184.1} \\
					& & & $4,5,5,6,6,7,7,8,8,9)$ & & & & \\
					& \multirow{2}{*}{$(0, 1/2, 1/2)$} & \multirow{2}{*}{7} & $(1,1,1,1,2,2,2,3,3,3,$ & \multirow{2}{*}{(-5.55,-4.33,1.79)} & \multirow{2}{*}{(2.26,2.05,1.79)} & \multirow{2}{*}{1243.9} & \multirow{2}{*}{923.7} \\
					& & & $4,5,5,6,6,7,8,8,9,9)$ & & & & \\
					\bottomrule
				\end{tabular}
			\end{footnotesize}
		\end{center}
		\caption{Example optimal designs are shown for the two example trial design scenarios (TDSs) given in Section~\ref{example}. For brevity, here $ENM(\tau)=ENM\{\tau|\boldsymbol{f},\boldsymbol{e},m,X(\boldsymbol{S})\}$. The boundary and expected number of measurement (ENM) values are given to two and one decimal place respectively.}\label{tab1}
	\end{table}
\end{footnotesize}

In all instances the ENM under the null and alternative hypotheses is substantially reduced compared to the corresponding fixed sample design. Specifically, in TDS1 it is by as much as 30.1\% under the null hypothesis, with $\boldsymbol{w}=(1/2,0,1/2)$, and 24.6\% under the alternative hypothesis, with $\boldsymbol{w}=(0,1/2,1/2)$.

For TDS1, we can observe that the ability to alter $\boldsymbol{X}$ has allowed the three designs to incorporate interim analyses without the need to increase the value of $m$. We also see that the optimal choice for $\boldsymbol{X}$ is the same for each value of $\boldsymbol{w}$. However, as would be expected, the ENMs required under the null and alternative hypotheses is strongly dependent on the choice of $\boldsymbol{w}$. For example, with $\boldsymbol{w}=(1/2,0,1/2)$ the value of $ENM(0)$ is lower, and the value of $ENM(\delta)$ higher, than with $\boldsymbol{w}=(1/3,1/3,1/3)$. A similar statement holds for the design with $\boldsymbol{w}=(0,1/2,1/2)$. This is a consequence of the identified optimal boundaries. For example, the design with $\boldsymbol{w}=(0,1/2,1/2)$ has $f_1=-5.05$, almost ensuring the trial will not stop early for futility, as no concern is given to stopping early under the null hypothesis.

Similar statements hold for TDS2. However, here, the optimal $\boldsymbol{S}$ varies for each considered $\boldsymbol{w}$. Additionally, in this case, setting $\boldsymbol{w}=(0,1/2,1/2)$ results in a design with approximately the same required sample size under the alternative hypothesis as for $\boldsymbol{w}=(1/3,1/3,1/3)$. It appears that this value is approximately the lowest that can be attained for this TDS. However, these two designs have noticably different features outside of $ENM(\delta)$. As for TDS1, $\boldsymbol{w}=(0,1/2,1/2)$ results in a design with lower values for the early efficacy stopping boundaries, and almost removes the possibility to stop early for futility. This is a feature that should be anticipated in general.

\subsection{Post-trial analysis}

\begin{figure}
	\centering
	\includegraphics[width= 6in]{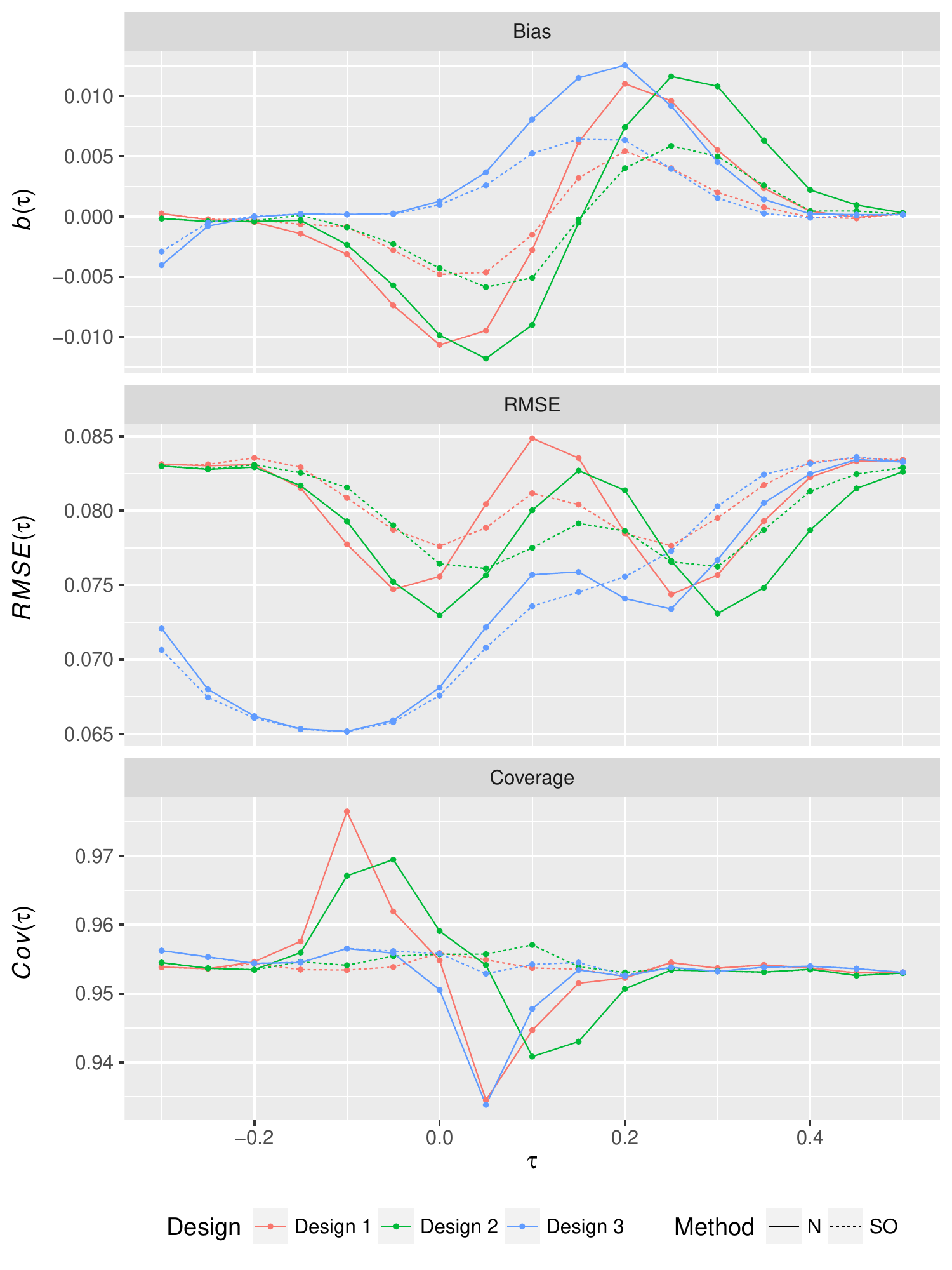}
	\caption{The estimated average bias and RMSE of the considered point estimates, and estimated coverage provided by the confidence intervals is shown from Trial Design Scenario 1. Design 1 is that shown in Table~\ref{tab1} for $\boldsymbol{w}=(1/3,1/3,1/3)$, Design 2 that for $\boldsymbol{w}=(1/2,0,1/2)$, and Design 3 that for $\boldsymbol{w}=(0,1/2,1/2)$. The naive (N) procedures are shown via the solid lines, and those based on the stagewise ordering (SO) via the dashed line.}\label{fig1}
\end{figure}

\begin{figure}
	\centering
	\includegraphics[width= 6in]{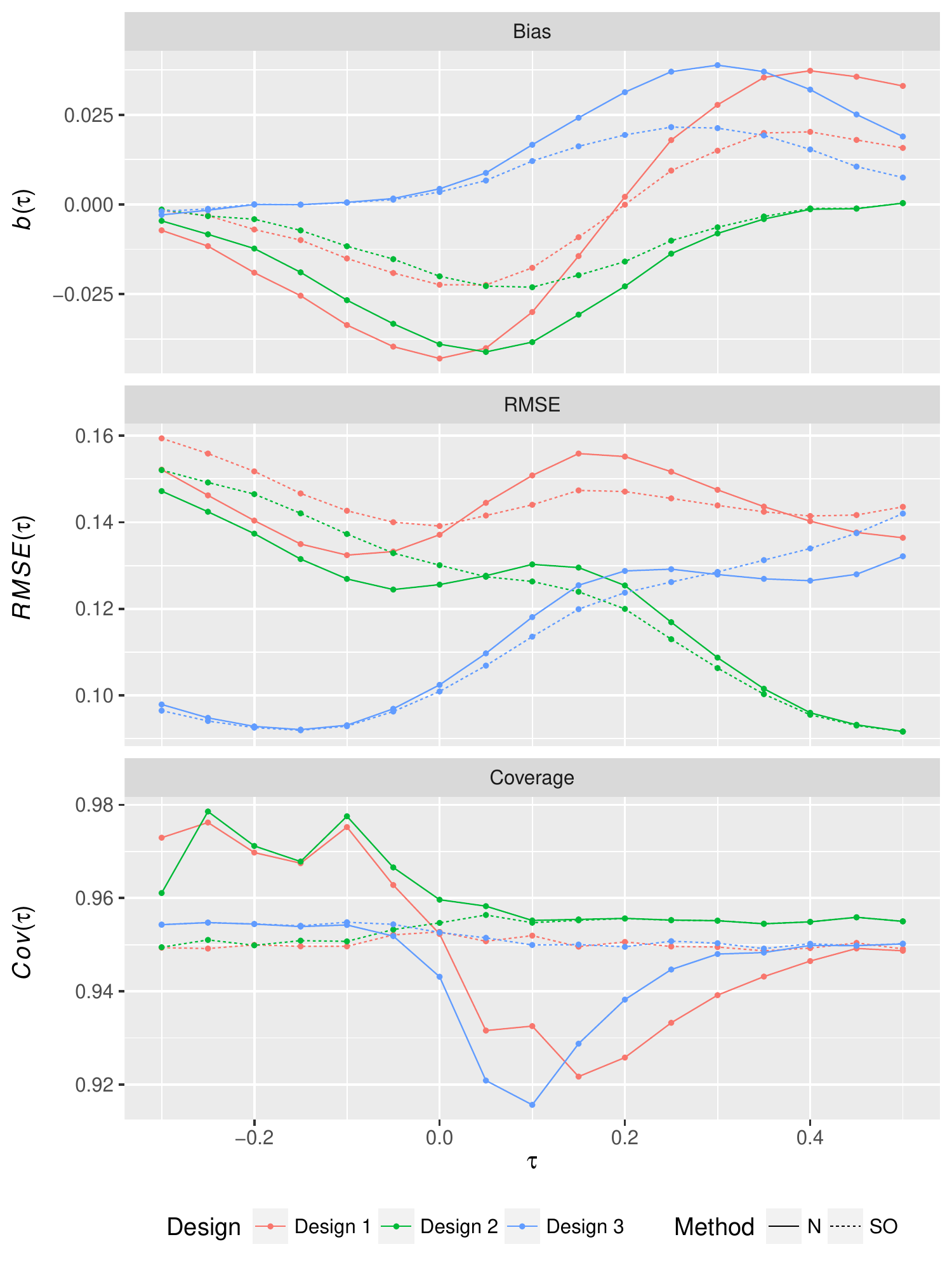}
	\caption{The estimated average bias and RMSE of the considered point estimates, and estimated coverage provided by the confidence intervals is shown from Trial Design Scenario 2. Design 1 is that shown in Table~\ref{tab1} for $\boldsymbol{w}=(1/3,1/3,1/3)$, Design 2 that for $\boldsymbol{w}=(1/2,0,1/2)$, and Design 3 that for $\boldsymbol{w}=(0,1/2,1/2)$. The naive (N) procedures are shown via the solid lines, and those based on the stagewise ordering (SO) via the dashed line.}\label{fig2}
\end{figure}

For the optimal designs given in Table~\ref{tab1}, a simulation study was conducted as described in Section~\ref{analysissec} to evaluate the performance of the naive and adjusted analysis procedures. In each case, we set $\mu=\pi_1=\dots=\pi_T=0$ for simplicity. Figure~\ref{fig1} displays the average bias and RMSE of the point estimates, as well as the coverage of the confidence intervals, in TDS1, when $\tau\in[-0.3,0.5]$. Figure~\ref{fig2} displays the equivalent findings for TDS2.

For both TDSs, the naive point estimator has, for certain values of $\tau$, large bias. The particular shape of the bias curve is largely dictated by the stopping boundaries. For example, in TDS2, the design with $\boldsymbol{w}=(1/2,0,1/2)$ (Design 2) will rarely stop early for efficacy given the optimal values of $e_1$ and $e_2$ in this instance. Accordingly, there is only ever a negative average bias in the point estimate. Similar statements hold for $\boldsymbol{w}=(0,1/2,1/2)$ (Design 3) in TDS2 also. These patterns are less clear in TDS1 however, as the optimal boundaries had a smaller magnitude on average. Importantly, the adjusted procedures almost universally reduce the average bias in the point estimate. However, this does come at a small cost to the RMSE in some instances. 

For the confidence intervals, the coverage of the naive procedures varies widely, reaching nearly 98\% for the design with $\boldsymbol{w}=(1/2,0,1/2)$ (Design 2), and below 92\% for $\boldsymbol{w}=(0,1/2,1/2)$ (Design 3), in TDS2 for certain values of $\tau$. In contrast, the adjusted method provides approximately the desired coverage across all designs, and all values of $\tau$.

\section{Discussion}\label{s:discussion}

In this paper, we have described methodology which allows the design of a group sequential SW-CRT to be optimised. In addition, we have detailed an approach through which post trial analysis can be conducted to account for the inclusion of interim analyses, reducing the bias in point estimates and improving the coverage over a naive final analysis. Ultimately, it is clear that our search procedure is able to identify highly efficient group sequential SW-CRT designs. In addition, the adjusted analysis procedures were able to substantially improve performance relative to the naive methods, reducing bias and providing coverage closer to the desired level.

One important point to discuss further is the influence of the choice of $\boldsymbol{w}$. It is clear from Table~\ref{tab1} that by altering $\boldsymbol{w}$ it is possible to improve performance, in terms of the ENM, under the null or alternative hypothesis. However, setting any of the $w_i$ equal to zero can lead to undesirable results. For example, $ENM(0)$ for the design with $\boldsymbol{w}=(0,1/2,1/2)$ in TDS1 was extremely large compared to the two other presented designs. This fact has been noted previously in other trial design settings \citep{mander}. It is advisable generally to set $w_i\neq0$ for $i=1,2,3$, and arguably the balanced optimal design with $\boldsymbol{w}=(1/3,1/3,1/3)$ should usually be preferred.

Another important result arose as a consequence of permitting the optimisation of $\boldsymbol{X}$. It is extremely common in group sequential trials that the maximal required sample size is larger than that of a corresponding fixed trial design. Here though, compared to a balanced fixed-sample SW-CRT, the optimal designs did not require any additional patients. This means that the optimal sequential designs can perform no worse than a balanced fixed-sample design in terms of the required number of measurements. Thus, as was discussed at length in Girling and Hemming (2016) \citep{girling} and Grayling et al. (2017c) \citep{grayling3}, if it is possible to utilise an optimised (typically non-balanced) $\boldsymbol{X}$, this should be done, as large efficiency gains will likely be possible. 

In summary, we have presented methodology to allow efficient group sequential SW-CRTs to be determined, and also to permit a corresponding post trial analysis to be performed with desirable performance. In practice, these methods are also easy to implement. We have provided code to replicate our presented results, but all that is required is the ability to evaluate multivariate normal integrals. One drawback that should be noted though is that there is no guarantee that the search procedure will converge to the truly optimal design; more likely it will only be approximately optimal. Ideally, one should therefore check convergence by performing multiple searches from different randomised starting values.

An interesting avenue for future work would be to examine whether there is any pattern to the optimal value of $\boldsymbol{X}$ depending on the choices for $\boldsymbol{w}$ and $\mathscr{T}$. In particular, whether these are similar to the findings obtained for fixed sample SW-CRTs would be of great interest, as this may then permit the run time of the search procedure for finding optimal sequential designs to be improved by utilising these results as the starting point.

Finally, it is important to recognise that the practical limitations of incorporating interim analyses in to a SW-CRT as discussed in Grayling et al. (2017a) \citep{grayling1} still stand. However, these notwithstanding, future trials should in our opinion consider incorporating at least one interim analysis in to a SW-CRT whenever possible. This will provide the majority of the efficiency gains from a sequential approach, whilst minimising the alterations required to the trials conduct.
	
\appendix

\section{Appendix}\label{appA}

\setcounter{table}{0}
\renewcommand{\thetable}{A.\arabic{table}}

\subsection{\texttt{CEoptim} initialisation}

In Section~\ref{optdes} we discussed our approach for simultaneously optimising $\boldsymbol{X}$, $\boldsymbol{f}$, $\boldsymbol{e}$, and $m$, when $C$ had been fixed. Ultimately, we were left to find the solution to the problem defined by Equation~(\ref{problem}), and proposed to use the function \texttt{CEoptim} in \texttt{R} for this. Here, we describe our initialisation of this function.

The problem~(\ref{problem}) imposes the following constraints on $\{(\boldsymbol{f},\boldsymbol{e}),m,\boldsymbol{S}\}$
\begin{enumerate}
	\item $m\in\mathbb{N}\backslash\mathbb{N}_1$.
	\item $S_c\in\mathbb{N}_{T+1}^+$ for $c\in\mathbb{N}_C^+$.
	\item $\min_{c\in\mathbb{N}_C^+}S_c \le t_1$.
	\item $\prod_{c\in\mathbb{N}_C^+}\mathbb{I}(S_c=t)=0$ for $t\in\mathbb{N}_T^+$.
	\item $\{\boldsymbol{f},\boldsymbol{e}\}\in\mathbb{R}^{|\mathscr{T}|}\times\mathbb{R}^{|\mathscr{T}|}$
	\item $f_i<e_i$ for $i\in\mathbb{N}_{|\mathscr{T}|-1}$.
	\item $f_{|\mathscr{T}|}=e_{|\mathscr{T}|}$.
\end{enumerate}
Here, constraints 2-4 are to ensure that $\boldsymbol{S}\in\mathscr{S}_{\mathscr{T}}$, whilst constraints 5-7 guarantee $\{\boldsymbol{f},\boldsymbol{e}\}\in\mathscr{B}_{|\mathscr{T}|}$.

In \texttt{CEoptim}, discrete parameters can be incorporated for by direct specification of a set of allowed values. Consequently, constraint 2 was included easily, whilst constraint 1 could be practically achieved by setting $m\in\mathbb{N}_{m_{\text{max}}}$ for some suitably large $m_{\text{max}}\in\mathbb{N}\backslash\mathbb{N}_1$. For our example, we set $m_{\text{max}}=10N_{SW}/CT$. Moreover, to ensure constraint 3 was adhered to, we designated without loss of generality that $S_1\in\mathbb{N}_{t_1}^+$.

To include constraints 5-7, rather than optimising the parameters $\boldsymbol{f}$ and $\boldsymbol{e}$, we transformed our problem to require optimisation of $\boldsymbol{f}$ and  $\boldsymbol{r}=(e_{1}-f_1,\dots,e_{|\mathscr{T}|-1}-f_{|\mathscr{T}|-1})^\top$. For, $\{\boldsymbol{f},\boldsymbol{r}\}$ completely defines a pair $\{\boldsymbol{f},\boldsymbol{e}\}$ conforming to constraints 5-7 if $\{\boldsymbol{f},\boldsymbol{r}\}\in\mathbb{R}^{|\mathscr{T}|}\times\{\mathbb{R}^+\}^{{|\mathscr{T}-1|}}$, which are linear constraints supported by \texttt{CEoptim}.

Unfortunately, constraint 4 could not be incorporated using the input parameters that can be passed to \texttt{CEoptim}. Nor would a convenient transformation of the problem work as above. Our solution therefore was, for each parameter set $\{(\boldsymbol{f},\boldsymbol{e}),m,\boldsymbol{S}\}$ considered in the search, to manually check whether constraint 4 was met. In the event that it was not, $O(\boldsymbol{f},\boldsymbol{e},m,\boldsymbol{S})$ was assigned to be infinite to ensure the design was rejected.

Note that in treating $m$ as discrete, we deviated from the approach taken in, for example, Wason (2015) \citep{wason3}. This was advantageous in that then only a single optimisation run was required to determine the optimal implementable design.

Now whilst the above ensured our determined optimal solution would confirm to all requirements placed on the design parameters, this is only one aspect of utilising \texttt{CEoptim}. Like many other global continuous optimisation routines, \texttt{CEoptim} allows the specification of several control parameters of the underlying algorithm's execution. These include an initial probability mass function for the discrete parameters, initial mean values and standard deviations for the continuous parameters, a sample size for each iteration of the algorithm, and a rarity parameter used in `elite' design selection. For more information on these, we refer the reader to Benham et al. (2017) \citep{benham}. 

Fortunately, the cross entropy method is relatively robust to its initialisation. Therefore we placed uniform distributions on the discrete parameters, as well as means of 0 and a (large) standard deviation of 10 on the continuous parameters. We also heeded the advice of Benham et al. (2017) \citep{benham}, and set the rarity parameter $\rho_{CE}=0.01$. Moreover, it was advised that the sample size for each iteration, $N_{CE}$, be chosen so that $N_{CE}\rho_{CE}$ is of the order 10 times the dimension of the optimisation problem. The dimension of our problem is $C+2|\mathscr{T}|$. Accordingly, we set $N_{CE}=10000(C+2|\mathscr{T}|)$. Of course, these choices may be sub-optimal, and attempts could be made to improve upon them. Practically however, the search procedure is robust and efficient enough that this is not necessary. Indeed, as was seen, for the considered examples our utilised approach led to efficient design determination. 

Finally, it is important to note the redundancy which exists in the specification of the vector $\boldsymbol{S}$, which we were forced to accept within our search procedure. This is a result of the fact that, for example, $O\{\boldsymbol{f},\boldsymbol{e},m,(1,2,3,4)^\top\}=O\{\boldsymbol{f},\boldsymbol{e},m,(2,1,3,4)^\top\}$. We may anticipate that we could remove the presence of this redundancy by enforcing $S_c\ge S_{c-1}$ for $c\in\mathbb{N}_C\backslash\mathbb{N}_2$, using either a manual check on candidate design proposal (as for constraint 4 above), or an approach similar to that taken for constraints 5-7 with the introduction of $\boldsymbol{r}$.

Unfortunately, the latter approach fails because there is then no effective means of ensuring that $S_c\in\mathbb{N}_{T+1}^+$ for $c\in\mathbb{N}_C^+$. Moreover, a manual check is only a viable solution when $C$ is small. To see this, consider the probability of randomly choosing an $\boldsymbol{S}$ with $S_c\ge S_{c-1}$ for $c\in\mathbb{N}_C\backslash\mathbb{N}_2$, which is given by

\begin{equation}\label{probcorr}
\sum_{s_1=1}^{t_1}\sum_{s_c=s_1}^{T+1}\dots\sum_{s_c=s_{c-1}}^{T+1}\prod_{c=1}^{C}\mathbb{P}(S_c=s_c).
\end{equation}

Suppose for example that $S_1 \sim U\{1,t_1\}$ and $S_c \sim U\{1,T+1\}$ for $c\in\mathbb{N}_C\backslash\mathbb{N}_2$, as is the case at the beginning of our search given the specifications above. Then, Table~\ref{tab2} presents the probability~(\ref{probcorr}) when $t_1=T/2$ for several $C\in\mathbb{N}\backslash\mathbb{N}_1$ and $T\in\mathbb{N}\backslash\mathbb{N}_1$. Clearly, without utilising an intractably large value for $N_{CE}$, this approach is in general not feasible as the chance of selecting a design that would pass our manual check is unacceptably small.

\begin{table}[htbp]
	\renewcommand{\arraystretch}{1.0}
	\begin{center}
			\begin{tabular}{rrrrrrrrrrr}
				\toprule
				 & \phantom{a} & $T=2$ & \phantom{a} & $T=4$ & \phantom{a} & $T=6$ & \phantom{a} & $T=8$ & \phantom{a} & $T=10$ \\
				\midrule
				$C=2$ && $1.0\times10^0$ && $0.9\times10^{-1}$ && $8.6\times10^{-1}$ && $8.3\times10^{-1}$ && $8.2\times10^{-1}$ \\
				$C=4$ && $3.7\times10^{-1}$ && $2.2\times10^{-1}$ && $1.7\times10^{-1}$ && $1.5\times10^{-1}$ && $1.3\times10^{-1}$ \\
				$C=6$ && $8.6\times10^{-2}$ && $2.9\times10^{-2}$ && $1.7\times10^{-2}$ && $1.2\times10^{-2}$ && $9.4\times10^{-3}$ \\
				$C=8$ && $1.6\times10^{-2}$ && $2.9\times10^{-3}$ && $1.1\times10^{-3}$ && $6.5\times10^{-4}$ && $4.4\times10^{-4}$ \\
				$C=10$ && $2.8\times10^{-3}$ && $2.4\times10^{-4}$ && $6.4\times10^{-5}$ && $2.8\times10^{-5}$ && $1.5\times10^{-5}$ \\
				$C=15$ && $2.5\times10^{-5}$ && $3.1\times10^{-7}$ && $2.6\times10^{-8}$ && $5.3\times10^{-9}$ && $1.7\times10^{-9}$ \\
				$C=20$ && $1.8\times10^{-7}$ && $2.7\times10^{-10}$ && $6.7\times10^{-12}$ && $5.7\times10^{-13}$ && $9.8\times10^{-14}$\\
				\bottomrule
			\end{tabular}
	\end{center}
	\caption{The probability of randomly generating a vector $\boldsymbol{S}=(S_1,\dots,S_C)^\top$ with $S_c\ge S_{c-1}$ for $c\in\mathbb{N}_C\backslash\mathbb{N}_2$, when $S_1 \sim U\{1,T/2\}$ and $S_c \sim U\{1,T+1\}$ for $c\in\mathbb{N}_C\backslash\mathbb{N}_2$, is shown for $(C,T)\in\{2,4,6,8,10,15,20\}\times\{2,4,6,8,10\}$.}\label{tab2}
\end{table}

\subsection{Simulation study}

Here, we describe how we generated trial data during the simulation study. The assumed correlation structure of the accrued responses in a SW-CRT means  more complex methods are required for this, compared to the simulation of individually randomised trials for example.

First, recall that $\boldsymbol{Y}_{t|m,\boldsymbol{X}}\sim N(\boldsymbol{D}_{t|m,\boldsymbol{X}}\boldsymbol{\beta},\boldsymbol{\Sigma}_{t|m,\boldsymbol{X}})$. For simplicity, from here drop explicit reference to $m$ and $\boldsymbol{X}$ and write $\boldsymbol{Y}_{t}\sim N(\boldsymbol{D}_{t}\boldsymbol{\beta},\boldsymbol{\Sigma}_{t})$. Then, for any set $\mathscr{P}=\{t_1,\dots,t_{|\mathscr{P}|}\}$ with $t_i\in\mathbb{N}_T^+$ for $i\in\mathbb{N}_{|\mathscr{P}|}^+$, define $\boldsymbol{Y}_{\mathscr{P}}=(\boldsymbol{Y}_{t_1}^\top,\dots,\boldsymbol{Y}_{t_{|\mathscr{P}|}}^\top)^\top$. Here for any $j\in\mathbb{N}_T^+$, $\boldsymbol{Y}_{j}$, $\text{dim}(\boldsymbol{Y}_{j})=mC\times1$, is the vector of responses from time period $j\in\mathbb{N}_T^+$.

Using this, write
\begin{equation*}
\begin{pmatrix} \boldsymbol{Y}_{\mathbb{N}_{t-1}^+} \\ \boldsymbol{Y}_{t} \end{pmatrix} \sim N\left\{ \begin{pmatrix} \boldsymbol{D}_{\mathbb{N}_{t-1}^+} \\ \boldsymbol{D}_{t} \end{pmatrix}\boldsymbol{\beta}, \begin{pmatrix} \boldsymbol{\Sigma}_{\mathbb{N}_{t-1}^+,\mathbb{N}_{t-1}^+} & \boldsymbol{\Sigma}_{\mathbb{N}_{t-1}^+,t} \\ \boldsymbol{\Sigma}_{t,\mathbb{N}_{t-1}^+} & \boldsymbol{\Sigma}_{t,t} \end{pmatrix} \right\},
\end{equation*}
with for any sets $\mathscr{P}$ and $\mathscr{Q}$, $\boldsymbol{D}_{\mathscr{P}}$ and $\boldsymbol{\Sigma}_{\mathscr{P},\mathscr{Q}}$ defined similarly as for $\boldsymbol{Y}_{\mathscr{P}}$ above.

Finally, we generate the $\boldsymbol{Y}_{t}$ as follows. Suppose that $\boldsymbol{Y}_{\mathbb{N}_{t-1}^+}$ has been generated for some $t\in\mathbb{N}_{T}\backslash\mathbb{N}_1$, $\boldsymbol{Y}_{\mathbb{N}_{t-1}^+}=\boldsymbol{y}_{\mathbb{N}_{t-1}^+}$ say. Then
\begin{equation*}
(\boldsymbol{Y}_t\mid\boldsymbol{Y}_{\mathbb{N}_{t-1}^+}=\boldsymbol{y}_{\mathbb{N}_{t-1}^+}) \sim N\{ \boldsymbol{D}_{t}\boldsymbol{\beta} + \boldsymbol{\Sigma}_{t,\mathbb{N}_{t-1}^+}\boldsymbol{\Sigma}_{\mathbb{N}_{t-1}^+,\mathbb{N}_{t-1}^+}^{-1}(\boldsymbol{y}_{\mathbb{N}_{t-1}^+}-\boldsymbol{D}_{\mathbb{N}_{t-1}^+}\boldsymbol{\beta}),\boldsymbol{\Sigma}_{t,t} - \boldsymbol{\Sigma}_{t,\mathbb{N}_{t-1}^+}\boldsymbol{\Sigma}_{\mathbb{N}_{t-1}^+,\mathbb{N}_{t-1}^+}^{-1}\boldsymbol{\Sigma}_{\mathbb{N}_{t-1}^+,1} \},
\end{equation*}
meaning $\boldsymbol{Y}_t$ can be generated using standard methods for generating random vectors following specified multivariate normal distributions.

\section*{Acknowledgements}

This work was supported by the Wellcome Trust [grant number 099770/Z/12/Z to M.J.G.]; and the Medical Research Council [grant number MC\_UU\_00002/3 to M.J.G. and A.P.M., and grant number MC\_UU\_00002/6 to D.S.R. and J.M.S.W.].

\end{document}